\documentclass[12pt]{iopart}
\usepackage{iopams}
\usepackage{graphics}
\usepackage{epsfig}
\usepackage{graphicx}
  
\begin{document}

\title[]{Expected Coalescence Rate of Double Neutron Stars for Ground Based Interferometers}
\author{T. Regimbau , J.A. de Freitas Pacheco, A. Spallicci and
  S. Vincent \footnote[3]{To whom correspondence should be addressed (regimbau@obs-nice.fr)}}

\address{Observatoire de la C\^ote d'Azur, BP 4229, 06304 Nice Cedex 4, France}

\begin{abstract}

In this paper we present new estimates of the coalescence rate of
neutron star binaries in the local universe and  we discuss its consequences
for the first generations of ground based interferometers. Our
approach based on both evolutionary and statistical methods gives a
galactic merging rate of 1.7 10$^{-5}$ yr$^{-1}$, in the range of previous
estimates 10$^{-6}$ - 10$^{-4}$ yr$^{-1}$. 
The local rate which includes
the contribution of elliptical galaxies is two times higher, in the
order of 3.4 10$^{-5}$ yr$^{-1}$.
We predict one detection every 148 and 125 years with initial VIRGO and LIGO, 
and up to 6 events per year with their advanced configuration. Our recent detection rate estimates from investigations on 
VIRGO future improvements are quoted.

\end{abstract}


\submitto{\CQG}

\maketitle

\section{Introduction}

The main uncertainties remain in estimating the detection rate (the

coalescence rate integrated over the volume of sensitivity of the
detector). For the past few years efforts have been made to calculate
the galactic coalescence rate, either from population synthesis (Potergies \& Spreeuw 1996; 
Portegies \& Yungelson 1998; Belczy\'nski et al. 2002) or from
statistical studies of the NS/NS population (Phinney 1991; van den
Heuvel \& Lorimer 1996; Kalogera et al. 2001; Kim, Kalogera \& Lorimer
2003; Kalogera et al. 2004). Both approaches have their own limitations: population synthesis have to deal with huge uncertainties on the parameters of the evolutionary model and statistical studies with the very small sample of observed NS/NS.
In this paper, we introduce a new method that combines, and takes
advantage, of the two methods. We first present the evolutionary model
from a pair of two massive stars to a neuton star binary, then we compute the galactic coalescence rate and the local rate and we derive the expected detection rate for the first two generations of ground based interferometers 
(VIRGO and few improved versions under investigation, initial and
advanced LIGO). Possible improvements in the detector sensitivity are also discussed.

\section{Formation of neutron star binaries}
We consider that NS/NS binaries are formed from pairs of massive stars according to the following scenario.
The most massive star of the system evolves rapidly into a red giant and explodes as a supernova, giving birth to a pulsar. If the binary system survives to the explosion, when the second star in its turn reaches the red giant stage and overflows its Roche Lobe, mass is transferred to the newly formed NS which is reaccelerated. The second star explodes as a supernova and if the system survives again, a NS-NS binary is formed. 
We assume that the youngest pulsars which don't undergo any mass
transfer from their companion have the same period evolution as
isolated pulsars but a kick velocity distribution close to that of
millisecond pulsars (Lyne \etal, 1998). Actually, because millisecond pulsars have been
reaccelerated by mass accretion in binary systems, the
supernova explosion of their progenitor was probably symmetric enough to keep
the system bounded. 

\section{The local coalescence rate}

\subsection{The galactic coalescence rate}
The galactic coalescence rate at the instant t can be expressed as:
\begin{equation}
\nu_c(t) = \lambda \beta_{ns} f_b \int_{\tau_0}^{(t-\tau_*-\tau_0)}R_*(t-\tau_*-\tau)P(\tau)d\tau 
\end{equation}
where
\begin{enumerate}
\item $R_*(t)$ is the star formation rate 
\item $\lambda$ the mass fraction of neutron star progenitors.\\
We assume a Salpeter mass distribution ranging  between 9-40 M$_{\odot}$ and
normalized between 0.1-80 M$_{\odot}$: 
\begin{equation}
\lambda = \int_9^{40}{\rm Am}^{-1.35}{\rm dm}\; {\rm with}\; {\rm A}=\int_{0.1}^{80}{\rm m}^{-2.35}{\rm dm}. 
\end{equation}
\item $f_b$ is the fraction of massive binaries formed among all stars
\item $\beta_{ns}$ is the fraction of massive binaries which remain bounded after the second supernova
\item $P(\tau)$ is the probability distribution for a newly formed
  neutron star binary to coalesce in a timescale $\tau$, $\tau_0$ is the minimum coalescence time and $\tau_*$ the mean timescale required for a newly formed massive binary to evolve into two neutron stars.
\end{enumerate}

\subsubsection{The galactic star formation rate}
The improvement of our method with respect to previous studies is that the galactic star formation rate is not taken proportional to the available mass of gas but it is reconstructed from observations of the chrosmospheric index for 552 stars. Figure 1, taken from the paper of Rocha-Pinto \etal (2000) shows the star formation rate with counting errors.
The enhanced periods of star formation around 1 Gyr, between 2-5 Gyr, between 7-9 Gyr are probably caused by accretion and merger episodes that make the disk grows and gains angular momentum (Peirani \etal, 2004).
\begin{figure}
\centering
\psfig{file=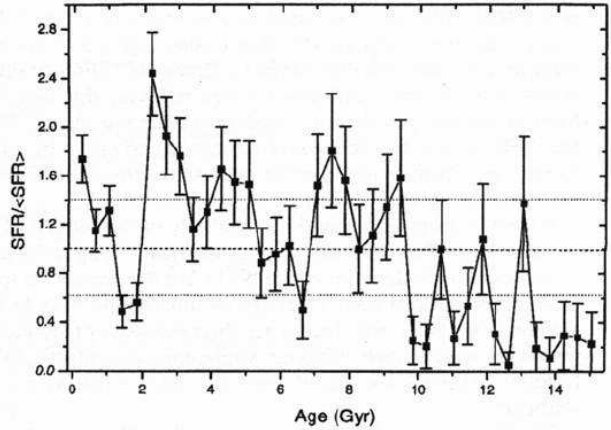,width=4in}
\caption{star formation rate with counting errors (Rocha-Pinto \etal,
  2000). The error bars correspond to an error of $\sqrt N$, where N is the number of stars found in each age bin and the dotted lines indicate the 2 sigma variations around the mean SFR}
\end{figure}

\subsubsection{Simulation of massive binary evolution}
\begin{figure}
\centering
\psfig{file=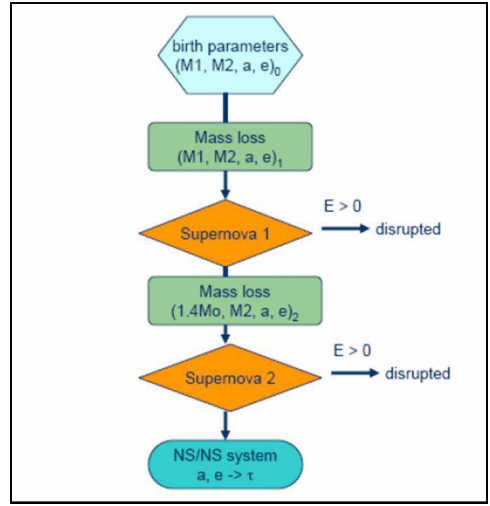,width=2.5in}
\caption{simulation pipeline of massive binary evolution}
\end{figure}
The fraction of massive binaries that remain bounded after the second
supernova explosion, the probability for a newly formed neutron star binary to
coalesce in a timescale $\tau$ as well as the minimum coalescence time are estimated from numerical simulations. 
For a given pair of massive NS the simulation code (figure 2) first computes the initial parameters (masses, orbital separation and eccentricity) and follows their evolution under mass transfer and mass loss due to stellar wind until the first supernova. 
The energy of the system is after calculated in order to determine
wether the system disrupts or not. If it remains bounded the same
procedure is followed until the second supernova and in the case a
double neutron star system is formed, the merging time is calculated from the orbital parameters.
After 500 000 numerical experiments we obtain a fraction
$\beta_{ns}=2.4\%$ of bound pairs with a probability distribution
$P(\tau)=0.087/ \tau$  to coalesce in a timescale $\tau$ ranging
between 2 10$^5$ yr and the age of the universe.

\subsubsection{Population synthesis of radio pulsars}

The fraction of massive binaries is derived from population synthesis
of isolated pulsars. We have used the method described in details in
Regimbau \& de Freitas Pacheco (2000), up-graded to take into account
the new pulsars discovered at high frequencies by the Parkes Multibeam
Survey as well as the most recent model for the kick velocity
distribution (Cordes \& Chernoff, 1997). The numerical code (figure 3) generates
population of pulsars with given sets of birth properties, follows
their evolution according to the dipole braking mechanism and model the
selection effects that constrain radio detection. The best agreement
between simulated and observed distributions of the period, the period
derivative and the distance determines the initial parameters of the
distributions of period and magnetic braking timescale (Table 1), as
well as the total number of pulsars (around 250000) and their
birthrate (one pulsar every 87 yr).
\begin{table}
\caption{birth distribution parameters of normal radio pulsars}
\centering
\lineup
\begin{tabular}{@{}ccc}
\br
 & mean & dispertion\\
\mr
 & $P_0 (ms) = 240 \pm 20$ & $\sigma_{P_0} = 80 \pm 20$\\
 & $ln \tau_0 (s) = 11 \pm 0.5$ & $\sigma_{ln\tau_0} = 3.6 \pm 0.2$\\
\br
\end{tabular}
\end{table}

\begin{figure}
\centering
\psfig{file=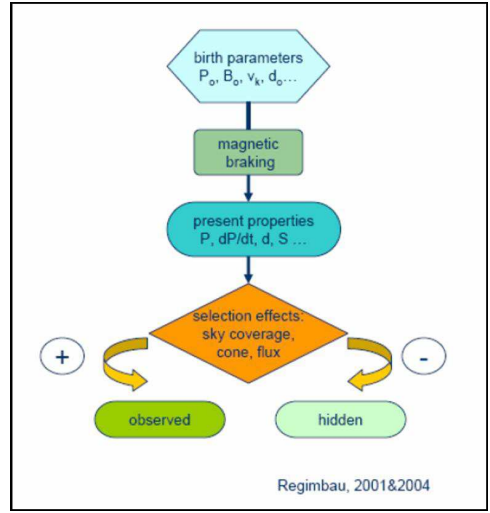,width=2.5in}
\caption{pipeline of population synthetis of normal radio pulsars}
\end{figure}
Our main assumption is that the youngest pulsars of the binary systems
have the same statistical birth properties as isolated pulsars except
for the kick velocity distribution that should be closer to that of
millisecond pulsars (Lyne \etal, 1998), follow the same magnetic dipole evolution period, and experience the same selection effects. 
Using the population synthesis code described before we find that a total number of ~730 pulsars is required to reproduce the observed sample of two young binary pulsars (PSR B1820-11 and PSR J0737-3039B). 
From the relation,
\begin{equation}
\frac{N_p}{N_b} = \frac{1}{\beta_{ns}}\frac{(1-f_b)}{f_b} + 2\frac{(1-\beta_{ns})}
{\beta_{ns}}
\end{equation}
we deduce $f_b = 0.136$.
Using the numbers obtained, it results for the present galactic
coalescence rate $\nu_S  = (1.7\pm 1.0)\times 10^{-5}$ yr$^{-1}$. The
estimated error is mostly due to uncertainties in the ratio $N_p/N_b$ derived from simulations.  
\subsection{The local coalescence rate}

Our local coalescence rate is the weighed average over spiral and
elliptical galaxies: 
\begin{equation}
\nu_c = \nu_S(f_S + f_E\frac{\nu_E}{\nu_S}\frac{L_S}{L_E})
\end{equation}
where $f_S=0.65$ and $f_E=0.35$ are the fractions of spiral and elliptical galaxies and $L_S$ and $L_E$ their luminosities. 
The fractions of formed massive stars and NS/NS binaries derived in the previous sections are assumed to be the same for all the galaxies. In revanche, the coalescence rate is taken to be equal to the galactic rate for spiral galaxies but is calculated separetely for ellipticals. Considing a star formation efficiency and an IMF derived from a recent model able to reproduce the observed color/magnitude diagram for 39 elliptical galaxies in VIRGO and COMA clusters (Idiart, Michard \& de Freitas Pacheco, 2003) we obtain a rate 
 $\nu_E$ = 8.6 10$^{-5}$ yr$^{-1}$ which is around five times higher than the Galaxy rate. 
\begin{figure}
\centering
\psfig{file=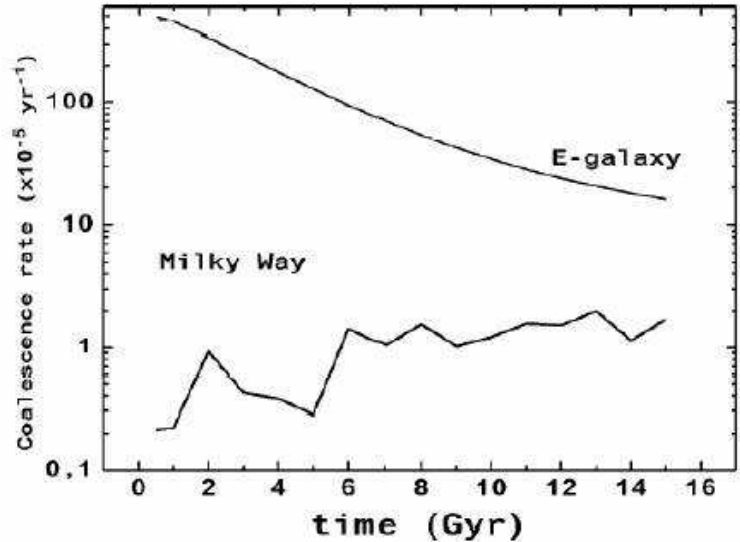,width=5in}
\caption{Evolution of coalescence rates for the Milk Way and for a typical E galaxy.
In ellipticals the bulk of stars is formed very early and the merging rate
reflects mainly the probability of a NS-NS binary to coalesce in a given timescale. For the
Galaxy, the star formation is continuous and intermittent, producing a modulation effect in
the coalescence rate history}
\end{figure}
Figure 4 compares the evolution of the coalescence rate for spiral and elliptical galaxies.
In ellipticals, stars are formed within the first 1-2 Myr and the curve behaviour corresponds essentially to the $1/\tau$  behavior of the merging time probability. 
\section{The detection rate}
For a given interferometer, the detection rate is obtained by scaling
the total luminosity within the volume probed by the detector with
respect to the luminosity of the Galaxy as: 
\begin{equation}
\nu(<D)=\nu_c \frac{L_V}{L_{MW}} \; {\rm with} \; V=\frac{4}{3} \pi D^{3}
\end{equation}
$L_v$ is estimated by real counts of galaxies using the LEDA catalog
(courtois \etal, 2004) which contains 106 galaxies and which completeness is about 84\% up to
B = 14.5. This procedure permits to account for the anisotropy of the
local universe (figure 5), in particular the huge concentration of
galaxies in the direction of the Norma/Centaurus cluster (the Great
Attractor).
\begin{figure}
\centering
\psfig{file=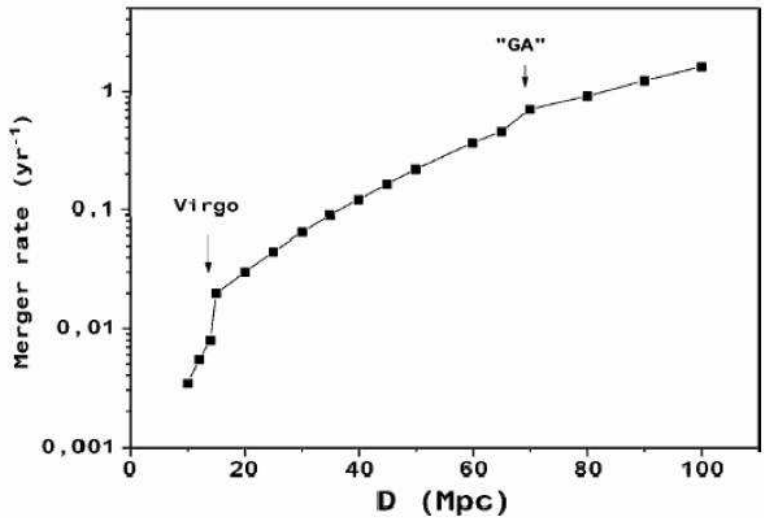,width=5in}
\caption{merging rate inside a volume $V=\frac{4}{3} \pi D^{3}$}
\end{figure}
Table 2 gives the distance probed by the interferometers VIRGO, LIGO and LIGO Ad assuming a S/N of 7 and a false alarm rate of 1, as well as the minimal detection rate required to have one detection.
Future developments of the detector configuration and data analysis techniques are expected to improve significantly the previous detection rates.

In a recent paper Spallicci \etal (2005) have shown that a detection rate of 1.5 yr$^{-1}$ could be reached by improving the initial VIRGO configuration in reducing either 
the overall noise by a factor of 10 in the mirror noise band (52-148 Hz) or the main noises by specific factors in the all bandwith.
An improvement by a factor of around 5-6 can also be obtained by
combining VIRGO with the two LIGO in a network of detectors (Pai,
Durandhar \& Bose). In this case the probed distance could be
increased up to 22 Mpc giving a detection rate of 1 event/26 yr.
\begin{table}
\caption{probed distance and event rate for initial VIRGO and LIGO and for advanced LIGO}
\centering
\lineup
\begin{tabular}{@{}ccc}
\br
VIRGO & LIGO & LIGO Ad\\
\mr
13 Mpc & 14 Mpc & 207 Mpc\\
1 event/148 yr & 1 event/125 yr & 6 events/yr\\
\br
\end{tabular}
\end{table}

\section{Conclusion}
In this paper the coalescence rate of neutron star binaries in the local universe is reviewed.
Our approach takes advantages on both evolutionary and statistical
methods and includes most recent observational results. We derive a
galactic merging rate of 1.7 10$^{-5}$ yr$^{-1}$, which is in the
range of previous estimates ($10^{-6}-10^{-4}$)  
The local rate which
includes the contribution of elliptical galaxies is two times higher,
in the order of  3.4 10$^{-5}$ yr$^{-1}$.
We predict one detection every 125 - 148 years with initial LIGO and
VIRGO, while advanced 
LIGO should be able to detect up to 
6 events per year.
In a next step, we plan to study the stochastic background created
by the superposition of all the double neutron star systems up to
redshift 5. For the first time numerical simulations based on monte
carlo techniques will be used to obtain a realistic estimate of the
spectrum of the energy density. This approach would permit to
understand better the statistics of the integrated signal and would be
very useful to develop adequate detection strategies.

\section*{References}
\begin{harvard}

\item[]Belczy\'nski K, Kalogera V and Bulik T 2002 ApJ {\bf 572} 407
\item[] Cordes J M and Chernoff D F 1997 ApJ {\bf 482} 971
\item[] Courtois H, Paturel G, Sousbie T and Labini F S 2004 A\&A {\bf
    423} 27
\item[] Idiart T P, Michard R and de Freitas Pacheco J A 2003 A\&A
  {\bf 398} 949
\item[] Kalogera V, Narayan R, Spergel D N and Taylor J H 2001 ApJ
  {\bf 556} 340
\item[] Kalogera V et al. 2002 ApJ {\bf 601} L179
\item[] Kalogera V et al. 2004 ApJ {\bf 614} L137
\item[] Kim C, Kalogera V and Lorimer D R 2003 ApJ {\bf 584} 985
\item[] Lyne A G, Manchester R N, Lorimer D R, Bailes M, D'Amico N,
Tauris T M, Johnston S, Bell J F and Nicastro L 1998 MNRAS {\bf 295}
\item[] Pai A, Dhurandhar S and Bose S 2002  {\it Phys. Rev. D} {\bf 64} 042004
\item[] Peirani S, Mohayaee R and de Freitas Pacheco J A 2004 MNRAS
  {\bf 348} 921
\item[] Phinney E S 1991 ApJ {\bf 380} L17
\item[] Potergies Zwart S F and Spreeuw H N  1996 A\&A {\bf 312} 670
\item[] Potergies Zwart S F and Yungelson L 1998 A\&A {\bf 332} 173
\item[] Regimbau T and de Freitas Pacheco J A 2001 A\&A {\bf 374} 182
\item[] Rocha-Pinto H J, Scalo J, Maciel W J and Flynn C 2000 ApJ {\bf
    31} L115
\item[] Spallicci A, Aoudia S, de Freitas Pacheco J A, Regimbau T, Frossati G
2005  \CQG  {\bf 22} S461.
\item[] van den Heuvel E P J and Lorimer D 1996 MNRAS {\bf 283} L37

\end{harvard}
\end{document}